\documentclass[submitting]{nst}

\usepackage{subfigure,dcolumn}
\usepackage[T2A,T1]{fontenc}
\usepackage[russian,english]{babel}

\usepackage{listings}
\usepackage{float}
\usepackage{tabularx}
\usepackage[normalem]{ulem}

\begin{document}

\title{
%
A phenomenological approach to direct ${\rm{K}}^{*}$ production and hadronic medium effects in nucleus–nucleus collisions at high baryon density
}

\author{Hongcan Li}
\affiliation{Key Laboratory of Quark and Lepton Physics (MOE) and Institute of Particle Physics, Central China Normal University, Wuhan 430079, China}

\author{Yun Liu}
\affiliation{Key Laboratory of Quark and Lepton Physics (MOE) and Institute of Particle Physics, Central China Normal University, Wuhan 430079, China}

\author{Guangyu Zheng}
\affiliation{School of Nuclear Science and Technology, University of Chinese Academy of Sciences, Beijing, 101408, China}

\author{Yaping Wang}
\thanks{Corresponding author: wangyaping@ccnu.edu.cn}
\affiliation{Key Laboratory of Quark and Lepton Physics (MOE) and Institute of Particle Physics, Central China Normal University, Wuhan 430079, China}

\author{Guannan Xie}
\affiliation{School of Nuclear Science and Technology, University of Chinese Academy of Sciences, Beijing, 101408, China}

\author{Gao-Chan Yong}
\affiliation{Institute of Modern Physics, Chinese Academy of Sciences, Lanzhou, 730000, China}

\begin{abstract}
Short-lived hadron resonances serve as sensitive probes of the late-stage hadronic medium in heavy-ion collisions.
Using the AMPT-HC model, we study ${\rm{K}}^{*}(892)$ production and its hadronic medium effects in Au+Au collisions at $\sqrt{s_{\rm{NN}}} = 3$ GeV, a region of high baryon density. 
We introduce a phenomenological direct-production mechanism for ${\rm K}^{*}$ by replacing a fraction of the final-state kaons produced in the ${\rm NN} \to {\rm NYK}$ and ${\rm MN} \to {\rm YK}$ channels with ${\rm K}^{*}$ resonances, with the substitution fraction controlled by a parameter $\alpha$ while conserving four-momentum.
T he direct ${\rm K}^{*}$ is produced early, at about 6 fm/$c$, with little centrality dependence, whereas resonance fusion via ${\rm K}+\pi\to{\rm K}^{*}$ occurs later, with the mean production time increasing from about 8 to 10 fm/$c$ toward central collisions. 
Consequently, direct ${\rm K}^{*}$ mesons suffer stronger daughter rescattering, leading to a pronounced decrease in reconstruction efficiency toward central collisions, while the ${\rm K}^{*}$ survival rate remains close to unity. 
Elastic scattering of the daughters also shifts the invariant mass away from the resonance peak, contributing to the background-like component. 
The ${\rm K}^{*}/{\rm K}$ centrality dependence reflects the competition between direct production and resonance fusion and is sensitive to $\alpha$. 
At 3 GeV, a moderate direct-production contribution may result in an increasing ${\rm K}^{*}/{\rm K}$ ratio toward central collisions, providing a testable prediction for future measurements.
\end{abstract}


\maketitle

\section{Introduction}
\label{sec:intro}

Hadron resonance states are important sources of particle production in heavy-ion collisions.
Depending on the theoretical framework, they can be formed through various mechanisms, such as string fragmentation, quark coalescence, or hadron scattering.
Among these, resonances produced by hadron scattering play a particularly significant role, especially in the few-GeV energy region.
For example, the $\pi + {\rm{N}}$ scattering cross section reaches about 200 mb at the $\Delta$ resonance peak~\cite{bib:PDG2026}.
Similarly, the ${\rm{K}} + \pi \to {\rm{K^*}}$ resonance scattering is estimated to be about 60 mb by Ko~\cite{bib:Ko1981}.
These sizable cross sections highlight the indispensable role of resonance production in determining the final particle yields, and they also influence the dynamical evolution of the medium and the final-state observables.

Among these short-lived resonances, the ${\rm{K}}^{*}(892)$ meson is of particular interest.
It has two charge states, the neutral ${\rm{K}}^{*0}$ and the charged ${\rm{K}}^{*+}$, with decay widths of 47.3 and 50.8 MeV, corresponding to lifetimes of $\sim 4.16$ and $\sim 3.87~\rm{fm}/c$, respectively~\cite{bib:PDG2026}.
Since these lifetimes are significantly shorter than the typical duration of a heavy-ion collision, which can extend to several tens of $\rm{fm}/c$ for Au+Au collisions, the ${\rm{K}}^{*}$ meson predominantly decays inside the medium.
Consequently, its decay daughters (${\rm{K}}$ and $\pi$) inevitably traverse the medium and undergo scattering before reaching the detectors~\cite{bib:Bleicher2002}.
Therefore, the ${\rm{K}}^{*}$ serves as a unique probe of the late-stage hadronic interactions in heavy-ion collisions.

The STAR Collaboration has systematically measured ${\rm{K}}^{*0}$ production at mid-rapidity in Au+Au collisions within the Beam Energy Scan (BES) program.
In the first phase of BES (BES-I), ${\rm{K}}^{*0}$ was measured at $\sqrt{s_{\rm{NN}}} = 7.7$--$39$ GeV, revealing a clear suppression of the ${\rm{K}}^{*0}/{\rm{K}}$ ratio in central collisions compared with peripheral collisions~\cite{bib:STAR2023}.
This suppression is significantly larger than predictions from thermal models that neglect final-state scattering.
To interpret this centrality dependence, two competing mechanisms have been considered: rescattering and regeneration.
Rescattering occurs when the decay daughters (${\rm{K}}$ and $\pi$) undergo elastic collisions with other hadrons in the medium, which alters their four-momenta and prevents the reconstruction of the parent ${\rm{K}}^{*}$, thereby reducing the measured yield.
Regeneration, on the other hand, could occur via the pseudo-elastic scattering ${\rm{K}} + \pi \to {\rm{K^*}}$, which would enhance the signal.
However, since the hadronic medium is populated with a large number of $\pi$ mesons and the decay daughters of ${\rm{K}}^{*}$ interact primarily with them, and since the $\pi-\pi$ scattering cross section is much larger than the $\pi-{\rm{K}}$ one, rescattering of the daughters is expected to strongly dominate over regeneration for ${\rm{K}}^{*}$.
Based on this framework, STAR attributes the observed suppression to the dominance of hadronic rescattering over regeneration, which naturally accounts for the reduced ${\rm{K}}^{*}/{\rm{K}}$ ratio in more central collisions; transport models such as UrQMD have been employed to further interpret these observations.
Building on this, the recently completed BES-II phase, with improved detector performance and significantly larger statistics, has confirmed and refined these findings: it demonstrates the same centrality-dependent suppression of the ${\rm{K}}^{*0}/{\rm{K}}$ ratio with higher statistical significance, and shows that the suppression is more pronounced at BES energies than at the highest RHIC and LHC energies within a given multiplicity bin~\cite{bib:STAR2026,bib:ALICE2015,bib:ALICE2016,bib:ALICE2017,bib:ALICE2020rescattering,bib:Kozlowski2024,bib:Chabane2026}. 
Despite these experimental and transport-model efforts, a microscopic understanding of how different production mechanisms of ${\rm{K}}^{*}$ contribute to the observed medium effects and the final ${\rm{K}}^{*}/{\rm{K}}$ ratio remains incomplete.

In this paper, we employ the AMPT-HC model to analyze the production mechanisms of ${\rm{K}}^{*}$ in detail and to systematically study how hadronic medium effects modify its yield.
The rest of this paper is organized as follows. The AMPT-HC model and the implementation of the ${\rm{K}}^{*}$ direct production mechanism are described in Sec.~\ref{sec:model}. The simulation results are presented and discussed in Sec.~\ref{sec:results}. A summary is given in Sec.~\ref{Summary}.

\section{Brief introduction of the model}
\label{sec:model}

\subsection{A Multi-phase transport model}
\label{sec:AMPT}
The AMPT model~\cite{bib:Lin2005} is widely used to study heavy-ion collisions at high energies, such as LHC and RHIC top energies. 
It includes simulation of partonic and hadronic phase which consist of four main components: parton production from nucleons by HIJING~\cite{bib:Gyulassy1994}, partonic interaction and transport by Zhang's parton cascade (ZPC)~\cite{bib:zhang1998}, hadronization by parton fragmentation or coalescence process, and hadronic interaction by a relativistic transport (ART)~\cite{bib:Li1995}.

The AMPT-HC model, a pure hadron cascade version, is designed to describe hadronic interactions in nucleus-nucleus collisions at center-of-mass energies of a few GeV. 
In this model, the parton production, partonic interaction, and hadronization processes are switched off.
The positions of the nucleons are initialized according to the Woods-Saxon nucleon density distribution, Eq.~(\ref{Wood-Saxon}):
\begin{gather}
    \label{Wood-Saxon}
    \rho(r)=\frac{\rho_{0}}{1+e^{-\frac{r-R}{d}}}
\end{gather}
where $R$ is the half-density radius, $d$ is the diffuseness parameter, and $\rho_{0}$ denotes the normal nuclear density.
The nucleon momenta are initialized with the local Thomas-Fermi approximation, Eq.~(\ref{Thomas-Fermi}):
\begin{gather}
    \label{Thomas-Fermi}
    p_{\rm F}(\rho)=0.197\times\left(\frac{3\pi^{2}\rho}{2}\right)^{1/3}
\end{gather}
which depends on the nuclear matter density $\rho$; the coefficient $0.197$ converts between units of $\rm fm^{-1}$ and GeV.
Subsequently, the evolution of these nucleons is simulated by solving the Hamiltonian canonical equations, Eq.~(\ref{HamiltonianEq}), in the presence of the fields they generate, including the electromagnetic field, kaon potentials~\cite{bib:Brown1994}, baryon mean fields~\cite{bib:Gale1990}, and others.
\begin{gather}
    \frac{d\vec{r}}{dt}=\frac{\partial H}{\partial \vec{p}} \notag \\
    \label{HamiltonianEq}
    \frac{d\vec{p}}{dt}=-\frac{\partial H}{\partial \vec{r}}
\end{gather}
The baryon mean-field, given by Eq.~(\ref{BaryonMeanField}), depends on the nuclear matter density $\rho$:
\begin{gather}
    U(\rho)=a\frac{\rho}{\rho_{0}}+b\left(\frac{\rho}{\rho_{0}}\right)^{\sigma} \notag \\
    a=\left(-29.81-46.9\frac{K+44.73}{K-166.32}\right)~\rm MeV \notag \\
    b=23.45\frac{K+255.78}{K-166.32}~\rm MeV \notag \\
    \sigma=\frac{K+44.73}{211.05}
    \label{BaryonMeanField}
\end{gather}
$K$ is the incompressibility parameter, which controls the stiffness of the baryon mean field and is set to 380 MeV in this work~\cite{bib:Li1995,bib:STAR2022,bib:Gao2025,bib:Li2026}. 
During the transport, particles---including mesons ($\pi$, $\rho$, $\omega$, $\eta$, $\rm{K}$, $\rm{K}^{*}$, $\phi$), baryons ($\rm{N}$, $\Delta$, $\rm{N}^{*}(1440)$, $\rm{N}^{*}(1535)$, $\Lambda$, $\Sigma$, $\Xi$, $\Omega$) and their anti-particles---can be produced or absorbed through inelastic collision channels.

\subsection{$\rm{K}^{*}$ production mechanism in AMPT model}
\label{sec:resonance}

In the original AMPT model, which can operate in either the default or the string melting (SM) version, $\rm{K}^{*}$ mesons are predominantly produced through string fragmentation and quark coalescence. 
In these partonic scenarios, a $\rm{K}^{*}$ is identified when the invariant mass of the produced quark and anti-quark pair lies close to the $\rm{K}^{*}$ mass pole~\cite{bib:Lin2005}. 
In the hadron cascade (HC) version, however, all partonic processes are switched off. 
Consequently, $\rm{K}^{*}$ production from string fragmentation and quark coalescence is entirely suppressed, and its production becomes exclusively driven by hadron-hadron scattering processes. 
The primary channels for $\rm{K}^{*}$ formation in the original AMPT-HC code are summarized in Table~\ref{tab:kstar_ampthc}.
Among these channels, the resonance fusion process ${\rm{K}} + \pi \to {\rm{K^*}}$ is the dominant source of $\rm{K}^{*}$ production, and its cross section is parameterized based on the Breit-Wigner formula~\cite{bib:Lopez2010,bib:ALICE2012,bib:Ko1981}. 
Note that this resonance fusion mechanism corresponds directly to the regeneration process discussed in the STAR experimental papers~\cite{bib:STAR2023,bib:STAR2026}. 
\begin{table}[h]
\centering
\caption{Summary of $\rm{K}^{*}$ production mechanisms in the original AMPT-HC hadronic cascade.}
\label{tab:kstar_ampthc}
\vspace{0.2cm}
\begin{tabular}{l l}
\hline
\textbf{Production process description} & 
\textbf{Channels} \\
\hline
Resonance fusion & 
${\rm{K}} + \pi \to {\rm{K^*}}$ \\
Meson-meson annihilation & 
$\pi + \rho/\omega \to  {\rm{K^*}} + {\rm{K}}$ \\
Vector meson exchange & 
${\rm{K}} + \rho/\omega(\phi) \to  {\rm{K^*}} + \pi(\rm{K})$ \\
\hline
\end{tabular}
\end{table}

\begin{figure}[ht]
  \centering
  \includegraphics[width=0.98\linewidth,clip]{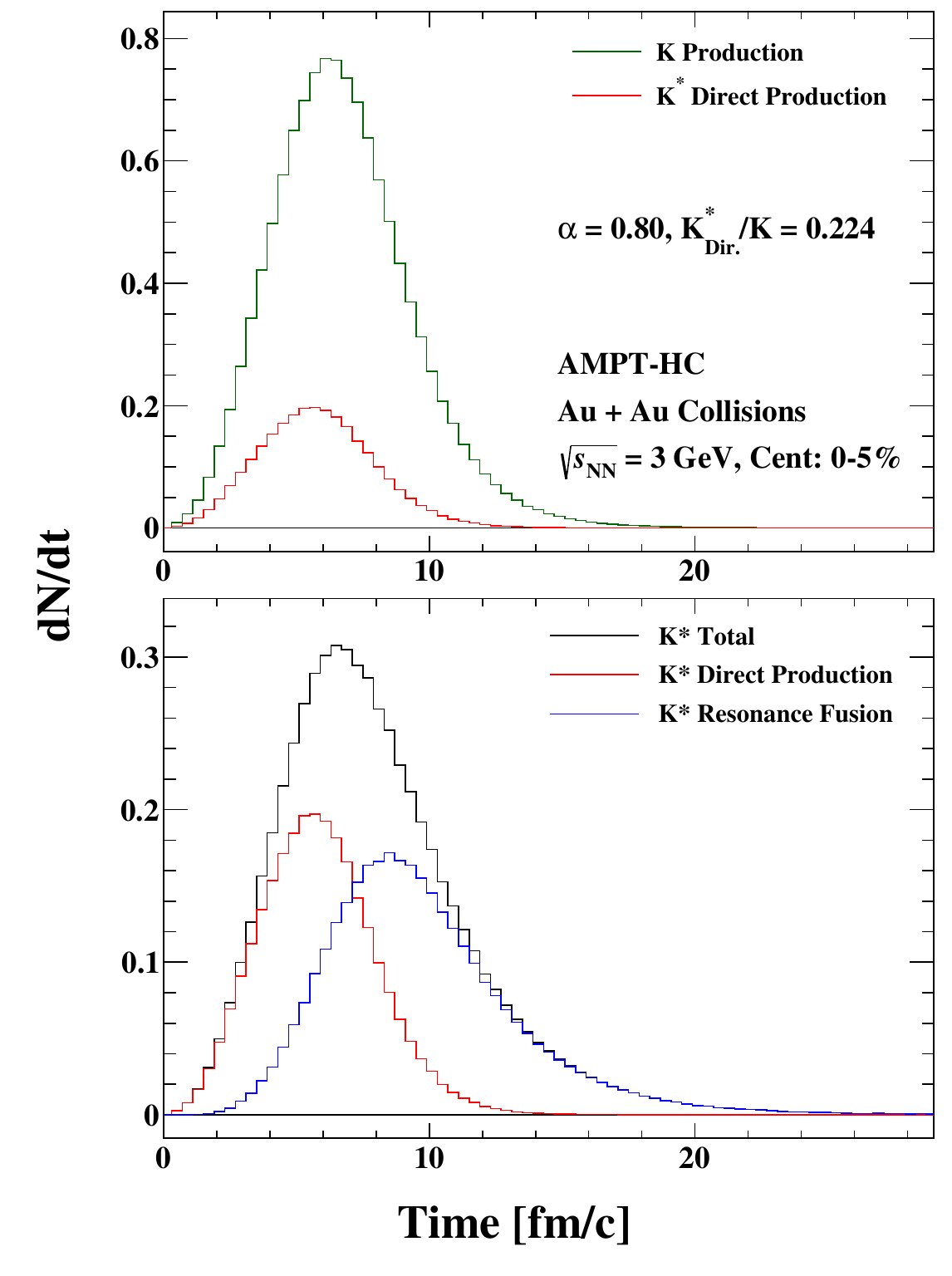}
  \caption{
  Time evolution of the production rate ($d{\rm{N}}/dt$) of ${\rm{K}}$ and ${\rm{K}}^{*}$ mesons in Au+Au collisions at $\sqrt{s_{\rm{NN}}}=3$ GeV with 0--5\% centrality from the AMPT-HC model with $\alpha = 0.8$.
  The upper panel shows the production rates of ${\rm{K}}^{+}$ (green line) and of ${\rm{K}}^{*}$ from the direct production mechanism (red line).
  The lower panel decomposes the total ${\rm{K}}^{*}$ production rate (black line) into the direct production (red line) and the resonance fusion production via ${\rm{K}} + \pi \to {\rm{K}}^{*}$ (blue line).
  The substitution ratio is $\alpha = 0.80$, corresponding to a direct ${\rm{K}}^{*}$ to kaon ratio of ${\rm{K}}^{*}_{\rm{Dir.}}/{\rm{K}} = 0.224$.
  }
  \label{fig:dndt}
\end{figure}
Nevertheless, in the default and SM versions, the quark and anti-quark pairs which form the $\rm{K}^{*}$ resonances originate from the primary ${\rm{NN}}$ collisions. 
When the partonic degrees of freedom are suppressed in the HC version, the loss of parton fragmentation and coalescence should not be taken to imply the complete disappearance of $\rm{K}^{*}$ production from these primary ${\rm{NN}}$ collisions. 
Instead, the $\rm{K}^{*}$ production mechanism naturally reverts to the hadronic level. 
Therefore, direct $\rm{K}^{*}$ production via primary ${\rm{NN}}$ scatterings is physically necessary~\cite{bib:HADES2015,bib:ALICE2012,bib:ALICE2020,bib:ALICE2020prc}. 
In addition, the nucleons inevitably interact with the hadronic medium created in heavy-ion collisions, which contains $\pi$, $\rho$, $\omega$, and other hadrons. 
These meson-nucleon ($\rm{MN}$) interactions serve as another essential source of $\rm{K}^{*}$ formation~\cite{bib:Napier1984}. 
However, a practical challenge arises because both experimental measurements and theoretical predictions for the $\rm{K}^{*}$ production cross section in ${\rm{NN}}$ and $\rm{MN}$ collisions are scarce. 
Given that the $\rm{K}^{*}$ resonance shares the same valence quark content as the $\rm{K}$ meson, we assume their hadronic production cross sections to be comparable. 
Thus, we model the direct $\rm{K}^{*}$ production by modifying the existing $\rm{K}$ production channels.

\begin{figure}[h]
  \centering
  \includegraphics[width=0.98\linewidth,clip]{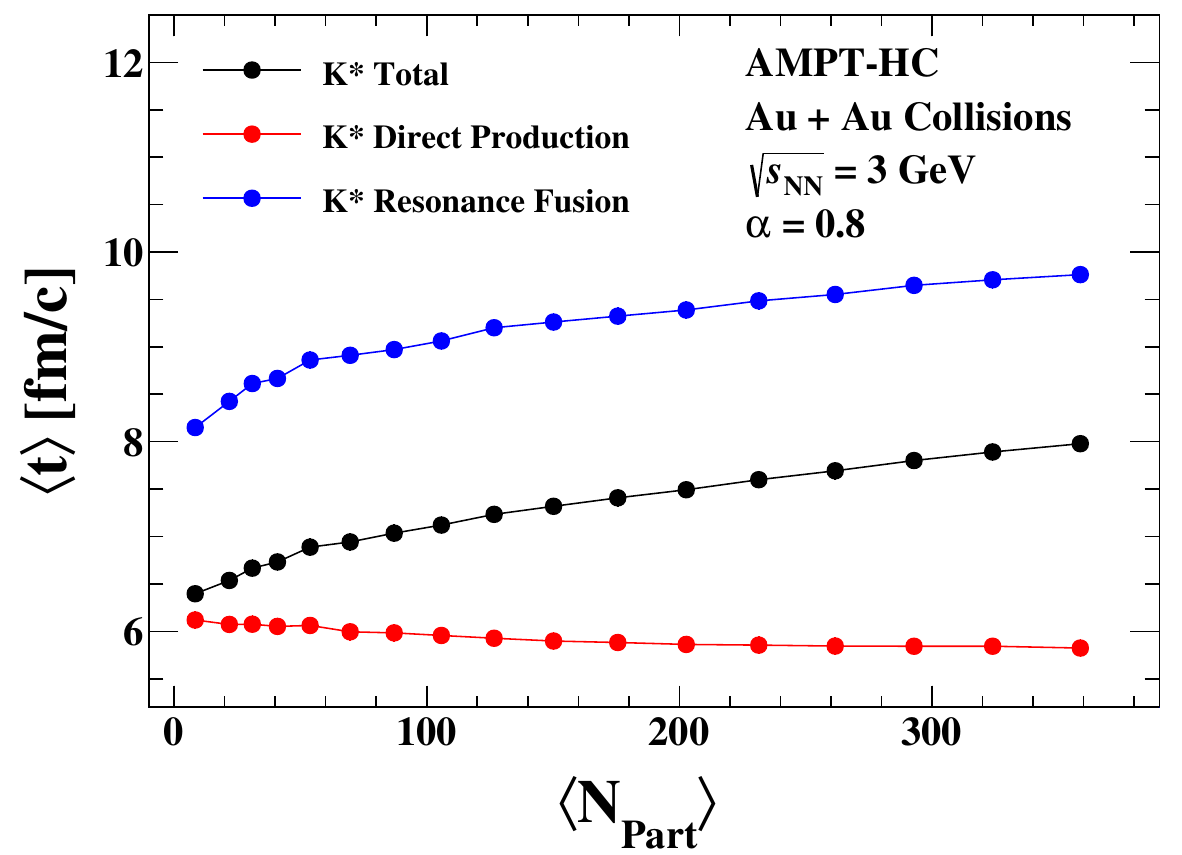}
  \caption{
  Mean production time $\langle t \rangle$ of ${\rm{K}}^{*}$ mesons as a function of the number of participating nucleons $\langle {\rm{N}}_{\rm{part}} \rangle$ in Au+Au collisions at $\sqrt{s_{\rm{NN}}} = 3$ GeV from the AMPT-HC model with $\alpha = 0.8$.
  The red, blue, and black lines represent ${\rm{K}}^{*}$ from the direct production mechanism, the resonance fusion mechanism, and all ${\rm{K}}^{*}$, respectively.
  }
  \label{fig:meantime}
\end{figure}
In the AMPT-HC model for $\sqrt{s_{\rm{NN}}}=3$ GeV Au+Au collisions, the primary sources of kaon production are the ${\rm{NN}} \to {\rm{NYK}}$ and ${\rm{MN}} \to {\rm{YK}}$ channels, where $\rm{N}$ represents a nucleon or its resonance state, $\rm{M}$ represents a $\pi$ or its resonance state, and $\rm{Y}$ represents a hyperon~\cite{bib:Zhang2026}. 
Therefore, our replacement algorithm is specifically designed to act on the final states of these two dominant reaction channels~\cite{bib:Fuchs1997}. 
We introduce a substitution ratio $\alpha$, representing the probability that the final-state $\rm{K}$ meson is replaced by a $\rm{K}^{*}$ resonance. 
Specifically, when a replacement is triggered, the mass of the produced $\rm{K}^{*}$ is sampled based on the Breit-Wigner distribution. 
The sampling range is constrained above the decay threshold $m_{\rm{K}^{*}} \ge m_{\rm{K}} + m_{\pi}$. 
Prior to the momentum reassignment, a kinematic check is performed to verify that the sum of the final-state masses does not exceed the available center-of-mass energy $\sqrt{s}$. 
If this condition is violated, the sampled mass is discarded and the original final state (with a $\rm{K}$ meson) is retained without further processing. 
As a result, the actual conversion rate is lower than the nominal $\alpha$ value, since a fraction of the sampled masses are discarded for violating energy conservation. 
Once the mass of \(\rm{K}^{*}\) is determined, the final-state momenta of all involved particles are recalculated to strictly satisfy four-momentum conservation. 
For the two-body process $\rm{MN \to YK^{*}}$, the momenta are directly derived from energy conservation in the center-of-mass frame; whereas for the three-body process $\rm{NN \to NYK^{*}}$, a modified multi-body phase-space generator is used. 
All calculated momenta are then rotated to align with the incident collision axis and boosted back to the laboratory frame via standard Lorentz transformations.

Fig.\ref{fig:dndt} shows the time evolution of the production rates ($d{\rm{N}}/dt$) for $\rm{K}$ and $\rm{K}^{*}$ mesons in Au+Au collisions at $\sqrt{s_{\rm{NN}}} = 3$ GeV with 0-5\% centrality, simulated by the AMPT-HC model with $\alpha = 0.8$. 
The upper panel shows the production rate of total kaons (green line) and of $\rm{K}^{*}$ from the direct production mechanism (red line). 
The lower panel further decomposes the total $\rm{K}^{*}$ (black line) into the direct production mechanism (red line) and the resonance fusion mechanism via ${\rm{K}} + \pi \to {\rm{K}}^{*}$ (blue line).

\begin{figure*}[ht!]
  \centering
  \includegraphics[width=0.98\linewidth,clip]{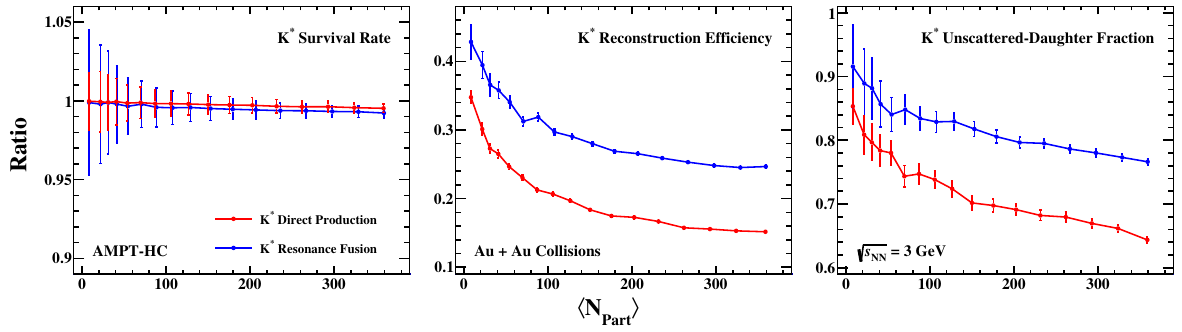}
  \caption{
  Hadronic medium effects on ${\rm{K}}^{*}$ mesons in Au+Au collisions at $\sqrt{s_{\rm{NN}}} = 3$ GeV, shown as a function of $\langle {\rm{N}}_{\rm{part}} \rangle$ for the direct production (red line) and the resonance fusion (blue line) mechanisms.
  The two mechanisms are obtained from two separate runs: the direct production from the run with the resonance-fusion channels switched off ($\alpha = 0.8$), and the resonance fusion from the $\alpha = 0$ run, in which all ${\rm{K}}^{*}$ originate from the resonance fusion mechanism.
  {\bf Left panel}: survival rate, the ratio of the number of ${\rm{K}}^{*}$ that decay to the number produced.
  {\bf Middle panel}: reconstruction efficiency, the ratio of the number of reconstructed ${\rm{K}}^{*}$ to the number of decays.
  {\bf Right panel}: fraction of reconstructed ${\rm{K}}^{*}$ whose daughter particles have not undergone elastic collisions before reconstruction.
  }
  \label{fig:threeratios}
\end{figure*}

\section{Results and discussion}
\label{sec:results}

In this section, we present the results obtained from the AMPT-HC model with the direct $\rm{K}^{*}$ production mechanism implemented.
We first compare the production timing of the two mechanisms and its centrality dependence, which determines how long the decay daughters spend inside the hadronic medium.
We then quantify how the hadronic medium modifies the $\rm{K}^{*}$ yield through the absorption and scattering of its decay daughters.
Finally, we study the ${\rm{K}}^{*}/{\rm{K}}$ ratio as the experimentally relevant observable and examine how the interplay of the two production mechanisms shapes its centrality dependence.

As shown in Fig.\ref{fig:dndt}, the ${\rm{K}}^{*}$ production under the two different mechanisms exhibits a clear sequential order. 
To quantitatively characterize this difference in formation timing, we compare their average production times $\langle t \rangle$ as a function of the number of participating nucleons $\langle {\rm{N}}_{\rm{part}} \rangle$, as shown in Fig.\ref{fig:meantime}. 
The red and blue lines represent the $\langle t \rangle$ of ${\rm{K}}^{*}$ from the direct production mechanism and the resonance fusion mechanism, respectively, while the black line represents all ${\rm{K}}^{*}$. 
A striking difference is observed between the two sources: the $\langle t \rangle$ of ${\rm{K}}^{*}$ from the direct production mechanism remains largely independent of centrality, staying constant at approximately \(6\ {\rm{fm}}/c\). 
In contrast, the ${\rm{K}}^{*}$ from the resonance fusion mechanism appears significantly later, with its $\langle t \rangle$ increasing gradually from roughly \(8\ {\rm{fm}}/c\) in peripheral collisions to nearly \(10\ {\rm{fm}}/c\) in the most central collisions. 
This sequential production has a clear physical origin. 
The direct ${\rm{K}}^{*}$, produced via the ${\rm{NN}} \to {\rm{NYK}}^{*}$ and ${\rm{MN}} \to {\rm{YK}}^{*}$ channels, originates from the early-stage nucleon-nucleon interactions and therefore appears at $\langle t \rangle \approx 6~\rm{fm}/c$ almost independently of centrality. 
In contrast, the resonance fusion process ${\rm{K}} + \pi \to {\rm{K}}^{*}$ is a secondary scattering process that requires the hadronic medium to first accumulate sufficient $\pi$ and ${\rm{K}}$ mesons; in denser, more central collisions this buildup takes longer, shifting the resonance fusion production to later times, with $\langle t \rangle$ increasing from $\sim 8$ to $\sim 10~\rm{fm}/c$. 
This timing difference has an important consequence for the medium effects discussed below: the daughters of the early-produced direct ${\rm{K}}^{*}$ spend a longer time inside the hadronic medium and are thus more likely to be absorbed or scattered.

\begin{figure}[h]
\centering
\includegraphics[width=0.98\linewidth,clip]{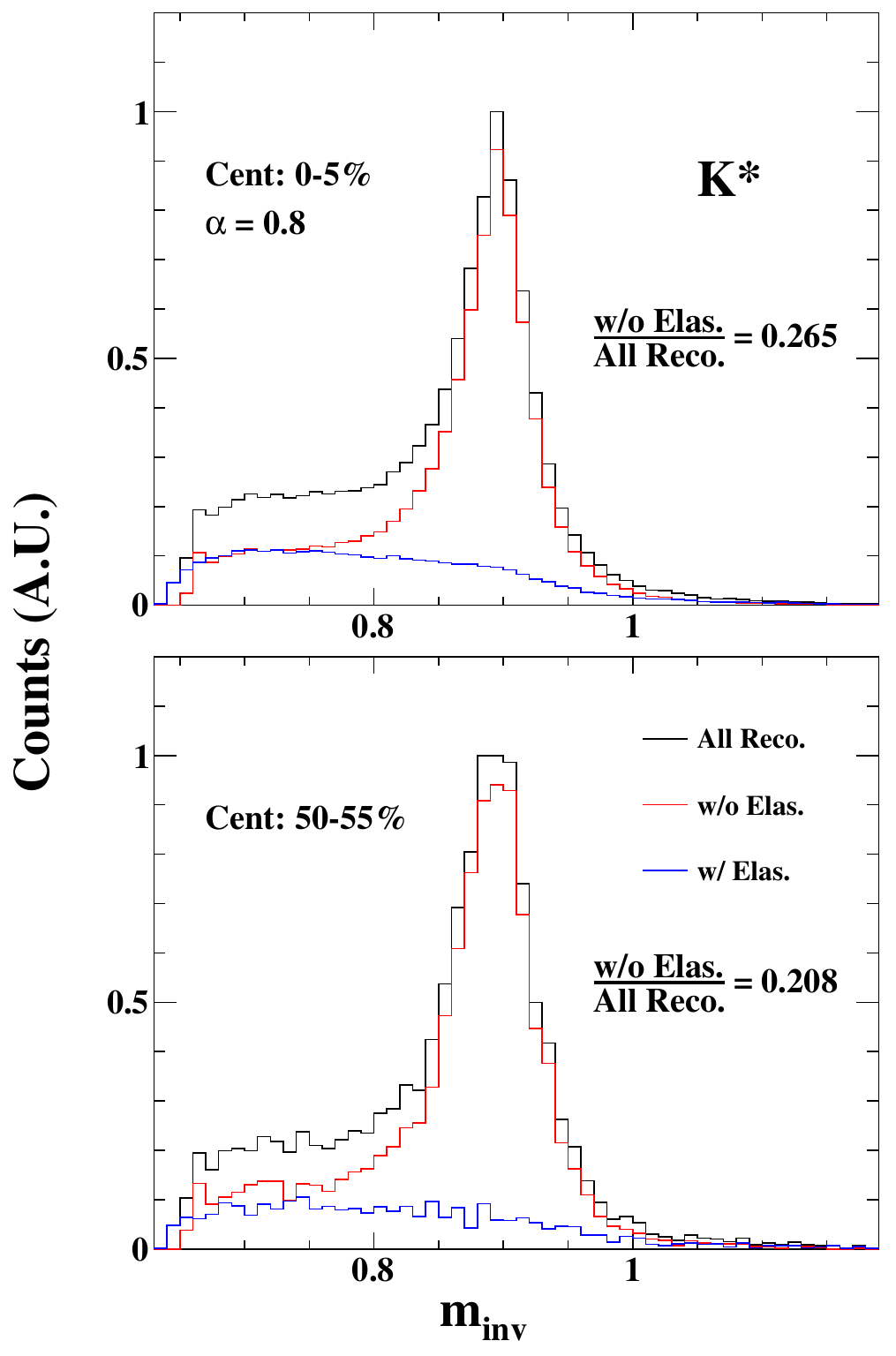}
\caption{
Invariant mass distributions of reconstructed ${\rm{K}}^{*}$ mesons in Au+Au collisions at $\sqrt{s_{\rm{NN}}} = 3$ GeV with $\alpha = 0.8$, for 0--5\% (top) and 50--55\% (bottom) centralities.
The black line represents all reconstructed ${\rm{K}}^{*}$, and the red and blue lines correspond to ${\rm{K}}^{*}$ whose decay daughters have not undergone and have undergone elastic collisions in the hadronic medium, respectively.
The fraction of reconstructed ${\rm{K}}^{*}$ with unscattered daughters is indicated in each panel.
}
\label{fig:minv}
\end{figure}

\begin{figure}[h!]
\centering
\includegraphics[width=0.98\linewidth,clip]{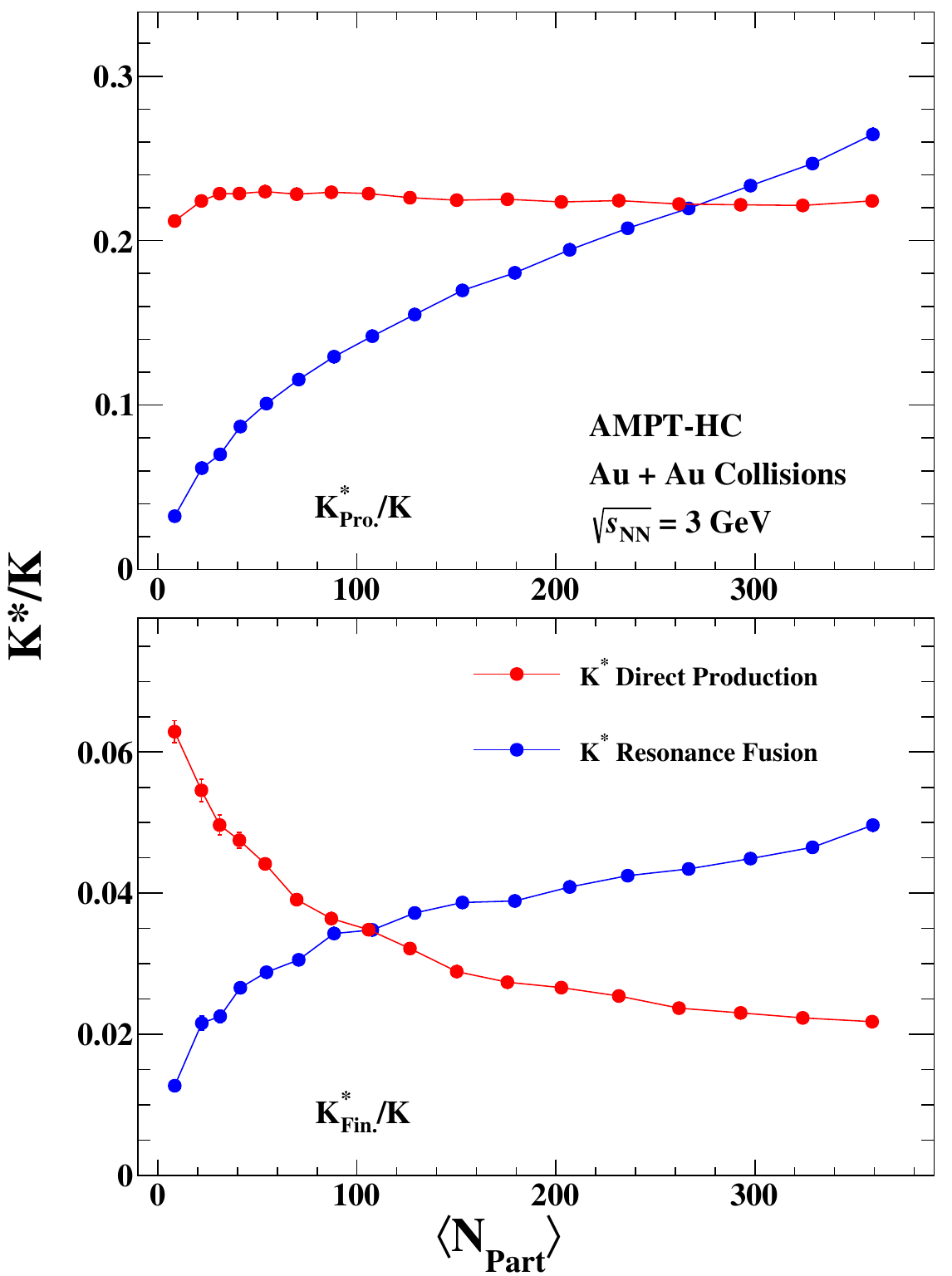}
\caption{
${\rm{K}}^{*}/{\rm{K}}$ ratio as a function of $\langle {\rm{N}}_{\rm{part}} \rangle$ in Au+Au collisions at $\sqrt{s_{\rm{NN}}} = 3$ GeV, decomposed into the direct production (red line) and the resonance fusion (blue line) contributions.
The two mechanisms are obtained from two separate runs: the direct production from the run with the resonance-fusion channels switched off ($\alpha = 0.8$), and the resonance fusion from the $\alpha = 0$ run.
{\bf Upper panel}: ratio of all produced ${\rm{K}}^{*}$ to ${\rm{K}}$.
{\bf Lower panel}: ratio of the final-state observable ${\rm{K}}^{*}$ to ${\rm{K}}$.
}
\label{fig:twotypesKstKratio}
\end{figure}

We now examine how the hadronic medium modifies the observable $\rm{K}^{*}$ yield. 
Fig.\ref{fig:threeratios} illustrates three distinct ratios as a function of $\langle {\rm{N}}_{\rm{part}} \rangle$ for both the direct production mechanism (red line) and the resonance fusion mechanism (blue line). 
The left panel displays the survival rate of $\rm{K}^{*}$, defined as the ratio of the number of $\rm{K}^{*}$ that decay to the total number produced. 
For both mechanisms, this ratio remains close to unity across all centralities, indicating that the absorption cross section of the ${\rm{K}}^{*}$ with the surrounding hadrons is small in the present model: the ${\rm{K}}^{*}$ mesons are rarely absorbed before they decay, and the yield loss of ${\rm{K}}^{*}$ is therefore dominated by the fate of its decay daughters rather than by the absorption of the parent resonance. 
This is precisely what makes ${\rm{K}}^{*}$ a sensitive probe of the hadronic medium: it decays inside the medium, and its daughters carry information about the medium they traverse. 
The middle panel presents the reconstruction efficiency of $\rm{K}^{*}$, defined as the ratio of the number of reconstructed $\rm{K}^{*}$ to the number of decays. 
The reconstruction efficiency decreases significantly from peripheral to central collisions, and the medium effects are stronger for the direct production mechanism: its efficiency drops from 0.37 to 0.15, a relative reduction of about 60\%, whereas that of the resonance fusion mechanism drops from 0.43 to 0.25, about 40\%. 
This pronounced drop indicates that the decay daughters of $\rm{K}^{*}$ have a high probability of interacting with and being absorbed by other hadrons in the medium; the stronger suppression of the direct production mechanism directly reflects its earlier production, since its daughters travel longer inside the medium and thus suffer more absorption and scattering. 
Consequently, these ${\rm{K}}^{*}$ cannot be reconstructed from the final-state particles and thus remain unobservable. 
The right panel shows the fraction of reconstructed $\rm{K}^{*}$ whose daughter particles have not undergone elastic collisions prior to reconstruction. 
This fraction also decreases from peripheral to central collisions. 
While these daughter particles are not absorbed, their four-momenta are altered by elastic collisions, which shifts the reconstructed invariant mass away from the true $\rm{K}^{*}$ mass and forms a mass distribution similar to the background. 
The impact of such elastic scattering on the invariant mass distributions is visualized in Fig.\ref{fig:minv}. 
The black, red, and blue lines represent all reconstructed $\rm{K}^{*}$, the reconstructed $\rm{K}^{*}$ whose daughters have not undergone elastic collisions, and the reconstructed $\rm{K}^{*}$ whose daughters have undergone elastic collisions, respectively. 
As clearly shown in the figure, the blue component no longer exhibits a Breit-Wigner resonance peak; instead, it forms a broad, flat distribution that resembles a background. 
In experimental analyses, such a distribution is typically subtracted as part of the background, and this background subtraction consequently leads to a further reduction in the observable $\rm{K}^{*}$ yield. 
This picture provides a microscopic realization of the rescattering scenario discussed by the STAR Collaboration, demonstrating on an event-by-event and daughter-by-daughter basis how rescattering translates into a reduction of the observable ${\rm{K}}^{*}/{\rm{K}}$ ratio.

\begin{figure*}[ht!]
\centering
\includegraphics[width=0.98\linewidth,clip]{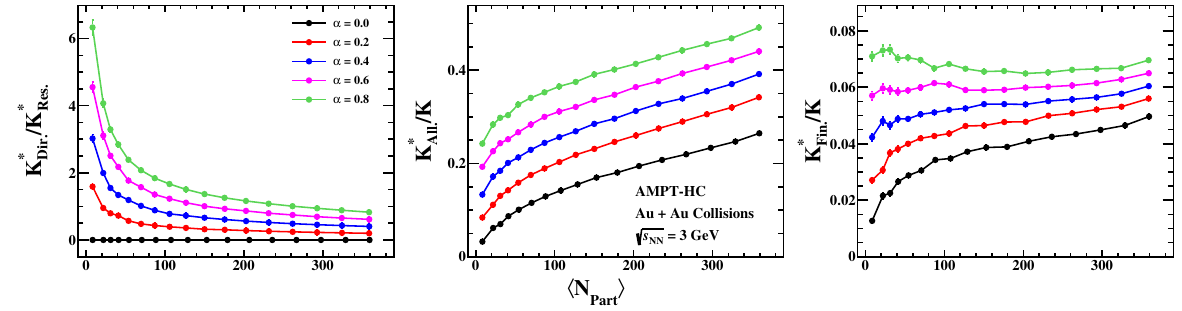}
\caption{
The ${\rm{K}}^{*}/{\rm{K}}$ ratio as a function of $\langle {\rm{N}}_{\rm{part}} \rangle$ for different substitution parameters $\alpha$, which controls the strength of the direct ${\rm{K}}^{*}$ production.
{\bf Left panel}: ratio of ${\rm{K}}^{*}$ from the direct production mechanism to that from the resonance fusion mechanism (${\rm{K}}^{*}_{\rm{Dir.}}/{\rm{K}}^{*}_{\rm{Res.}}$).
{\bf Middle panel}: ratio of all produced ${\rm{K}}^{*}$ to ${\rm{K}}$ (${\rm{K}}^{*}_{\rm{All.}}/{\rm{K}}$).
{\bf Right panel}: ratio of the final-state observable ${\rm{K}}^{*}$ to ${\rm{K}}$ (${\rm{K}}^{*}_{\rm{Fin.}}/{\rm{K}}$) after hadronic medium effects.
Different colors correspond to $\alpha = 0.0$, $0.2$, $0.4$, $0.6$, and $0.8$.
}
\label{fig:KstKratio_Alpha}
\end{figure*}

We now investigate how the two different production mechanisms contribute to the ${\rm{K}}^{*}/{\rm{K}}$ ratio. 
Fig.\ref{fig:twotypesKstKratio} separates the ${\rm{K}}^{*}/{\rm{K}}$ ratio into contributions from the direct production mechanism (red line) and the resonance fusion mechanism (blue line). 
In the upper panel, all ${\rm{K}}^{*}$ produced during the entire heavy-ion collision are used to evaluate the ${\rm{K}}^{*}/{\rm{K}}$ ratio without hadronic medium effects. 
As can be seen, the ${\rm{K}}^{*}_{\rm{Dir.}}/{\rm{K}}$ ratio remains nearly flat across all collision centralities, which reflects the similar production channels of the direct ${\rm{K}}^{*}$ and ${\rm{K}}$. 
In contrast, the ${\rm{K}}^{*}_{\rm{Res.}}/{\rm{K}}$ ratio exhibits a pronounced increase with $\langle {\rm{N}}_{\rm{part}} \rangle$, reflecting the fact that the ${\rm{K}}~+~\pi \to {\rm{K}}^{*}$ regeneration process is enhanced in denser media. 
In the lower panel, the final-state observable ${\rm{K}}^{*}$ is used to evaluate the ${\rm{K}}^{*}/{\rm{K}}$ ratio with hadronic medium effects. 
Here, a striking change occurs: the ${\rm{K}}^{*}_{\rm{Dir.}}/{\rm{K}}$ ratio shows a strong decreasing trend with centrality, while the ${\rm{K}}^{*}_{\rm{Res.}}/{\rm{K}}$ ratio still increases with centrality.

The overall ${\rm{K}}^{*}/{\rm{K}}$ ratio is determined by the combination of these two production mechanisms and depends on their relative yields. 
To quantitatively illustrate the interplay between them, we investigate the dependence on the substitution ratio $\alpha$, which governs the relative strength of the direct production mechanism. 
Fig.\ref{fig:KstKratio_Alpha} presents the ${\rm{K}}^{*}/{\rm{K}}$ ratio as a function of $\langle {\rm{N}}_{\rm{part}} \rangle$ for different $\alpha$ values. 
The left panel shows the ${\rm{K}}^{*}$ ratio of the direct production mechanism to the resonance fusion mechanism (${\rm{K}}^{*}_{\rm{Dir.}}/{\rm{K}}^{*}_{\rm{Res.}}$). 
For any given $\alpha>0$, this ratio decreases significantly with increasing centrality, indicating that the resonance fusion mechanism becomes increasingly dominant in more central collisions. 
The middle panel displays the ratio for all produced $\rm{K}^{*}$ mesons. 
As expected, increasing $\alpha$ leads to a higher overall production ratio across all centralities, since a larger fraction of $\rm{K}$ mesons is replaced by $\rm{K}^{*}$ at the production stage. 
The right panel displays the final observable ratio after the hadronic medium effects are included. 
For $\alpha=0$ (black line), where the direct production mechanism is absent, the final ratio exhibits a clear rising trend with centrality, driven purely by the late-stage resonance fusion mechanism. 
As $\alpha$ increases, the centrality-dependent suppression arising from the direct production component gradually takes over.

To summarize the physics picture, the direct production mechanism, originating from the early stage of the collision, drives the decreasing trend of ${\rm{K}}^{*}/{\rm{K}}$ with centrality because its decay daughters suffer strong medium effects; the resonance fusion mechanism, operating at later times, produces an increasing trend because it is enhanced in denser media. 
The measured ${\rm{K}}^{*}/{\rm{K}}$ ratio reflects the competition between these two contributions.
The centrality dependence of ${\rm{K}}^{*}/{\rm{K}}$ at $\sqrt{s_{\rm{NN}}} = 3$ GeV may differ qualitatively from that observed at higher energies. 
At $\sqrt{s_{\rm{NN}}} = 7.7$ GeV and above, the STAR data exhibit a decreasing ${\rm{K}}^{*}/{\rm{K}}$ ratio with centrality, attributed to the dominance of hadronic rescattering over regeneration. 
Whether this decreasing trend persists at 3 GeV, however, depends on the relative strength of the direct production mechanism. 
If the substitution parameter $\alpha$ is small, the resonance fusion mechanism dominates and the ${\rm{K}}^{*}/{\rm{K}}$ ratio would instead rise with centrality (Fig.~\ref{fig:KstKratio_Alpha}, right panel). 
The STAR measurement of ${\rm{K}}^{*}/{\rm{K}} \sim 0.2$ at 7.7 GeV provides a useful reference for the magnitude of $\alpha$: such a value can already be reproduced with a relatively small substitution parameter ($\alpha \sim 0.2$; see Fig.~\ref{fig:KstKratio_Alpha}, middle panel). 
With such a moderate $\alpha$, the resonance fusion mechanism dominates, and the ${\rm{K}}^{*}/{\rm{K}}$ ratio exhibits a rising trend with centrality. 
The rising scenario therefore appears quite plausible at 3 GeV, in contrast to the decreasing trend observed at higher energies; this can be tested by future measurements in the high baryon density region.


\section{Summary}
\label{Summary}

In summary, we have implemented a direct production mechanism for ${\rm{K}}^{*}$ mesons in the AMPT-HC model by substituting a fraction of the final-state kaons in the dominant ${\rm{NN}} \to {\rm{NYK}}$ and ${\rm{MN}} \to {\rm{YK}}$ channels with ${\rm{K}}^{*}$ resonances, controlled by a substitution parameter $\alpha$. 
Owing to the limited center-of-mass energy at $\sqrt{s_{\rm{NN}}} = 3$ GeV, the actual conversion rate is considerably lower than the nominal $\alpha$, since a fraction of the sampled masses is discarded by the energy-conservation check.

Using this model, we have systematically studied the production and the hadronic medium effects of ${\rm{K}}^{*}$ in Au+Au collisions at $\sqrt{s_{\rm{NN}}} = 3$ GeV. 
The direct ${\rm{K}}^{*}$ is produced early, at $\langle t \rangle \approx 6$ fm/$c$, almost independently of centrality, whereas the resonance fusion production ${\rm{K}} + \pi \to {\rm{K}}^{*}$ occurs later, with its mean production time increasing from about 8 to 10 fm/$c$ towards central collisions. 
As a consequence, the decay daughters of the direct ${\rm{K}}^{*}$ spend a longer time inside the hadronic medium and thus suffer stronger absorption and elastic scattering. 
The survival rate of ${\rm{K}}^{*}$ remains close to unity, indicating that the absorption of the parent resonance is weak; however, the reconstruction efficiency and the fraction of reconstructed ${\rm{K}}^{*}$ with unscattered daughters both decrease towards central collisions. 
In particular, the elastic scattering of the daughters shifts the reconstructed invariant mass away from the resonance peak, producing a broad distribution that is subtracted as background in experimental analyses, which further reduces the observable ${\rm{K}}^{*}$ yield, with a stronger reduction for the direct production mechanism.

The centrality dependence of the ${\rm{K}}^{*}/{\rm{K}}$ ratio is shaped by the competition between the two mechanisms and depends on the substitution parameter $\alpha$. 
Our results suggest that, at the high baryon density of 3 GeV, this dependence may differ from the decreasing trend observed at higher energies, a prediction that can be tested by future measurements.

\section*{Acknowledgements}
This work is supported in part by the National Natural Science Foundation of China under Grant No. 12375134 and No. 12305146, the National Key Research and Development Program of China (Grant No. 2024YFE0110103 and 2024YFA1611003), the Fundamental Research Funds for the Central Universities (Grant No. CCNU25JCPT017), and the CAS Project for Young Scientists in Basic Research (Grant No. YSBR-088).


\end{document}